# Role of interface in ferromagnetism of (Zn,Co)O films

M. Godlewski[1,2], E. Guziewicz[1], M.I. Łukasiewicz[1], I.A. Kowalik[1], M. Sawicki[1], B.S. Witkowski[1], R. Jakieła[1], W. Lisowski[3], J.W. Sobczak[3], M. Krawczyk[3]

[1] *Institute of Physics PAS, Al. Lotników 32/46, 02-668 Warsaw, Poland*
[2] *Dept. Mathematics and Natural Sciences, College of Sciences UKSW, Dewajtis 5, 01-815 Warsaw, Poland*
[3] *Institute of Physical Chemistry, PAS, ul. Kasprzaka 44/52, 01-224 Warsaw, Poland*

We demonstrate that room temperature ferromagnetic response (RT FR) of ZnCoO films grown at low temperature by the Atomic layer Deposition (ALD) method is due to Co metal accumulations at the ZnCoO/substrate interface region. The accumulated experimental evi evidences allow us to reject several other explanations of this effect in our samples, despite the fact that some of them are likely to be responsible for the low temperature FM in this class of the material.

## 1. Introduction

ZnCoO is the most studied oxide from the family of diluted magnetic semiconductors (DMS). There are two reasons for this. First, ZnCoO was the first DMS material for which room temperature (RT) ferromagnetism (FM) was reported [1]. Second, Co magnetic moment per ion is the largest among other transition metal (TM) ions in ZnO [2].

On the other hand, despite enormous experimental effort worldwide the RT FR in ZnCoO is far from being understood. Reports from different groups contradict each other and reproducibility of the results is low [1,3]. Actually our results shown in Fig. 3 demonstrate the similar situation. However, it has became clear that the uniformity of transition metal (TM) distribution is the most important factor influencing magnetic properties, as we have demonstrated in the case of ZnMnO [4-5].

Some of the present authors have already demonstrated that highly uniform ZnMnO films are paramagnetic [4], whereas those with nonuniform Mn distribution show ferromagnetic response either at low temperature or at RT. High uniformity of Mn distribution in ALD-grown (ALD stands for atomic layer deposition) layers was achieved by applying special sequence of the precursors cycles and lowering the growth temperature to $160^{o}C$ [4-5].

Recently we have shown that it is possible to obtain highly uniform ZnCoO films by the ALD technique [6]. For this the growth temperature has to be lowered below 200°C and the ZnO to CoO ratio of the ALD cycles should be around 8:1. Such layers show paramagnetic properties down to very low temperatures independently of sample thickness [6]. In the present study we discuss properties of nonuniform ZnCoO films. Such films were either grown at different growth modes (ratio of ZnO to CoO cyles was changed from 8:1 to 2:1) or we used a post-growth annealing of uniform samples. We discuss the origin of RT FR observed for these films. Some data for uniform samples are also given for the comparison.

## 2. Samples

All studied ZnCoO films were grown by the ALD on silicon substrates using DEZn, $Zn(acac)_2$, $H_2O$ and Cobalt(II) acetylacetonate as Zn, O and Co precursors, respectively. We used two different methods to obtain films with a nonuniform Co distribution. Firstly, a modified growth protocol was used, namely the ZnO to CoO pulse ratio was reduced to 2:1 and the growth temperature was increased (up to 300°C). The nonuniform samples were obtained by a post-growth annealing of the uniform ZnCoO films (annealing temperature up to 800°C). Further details on growth conditions can be found in [7]. Uniformity of ZnCoO samples was assessed by the Secondary Ion Mass Spectroscopy (SIMS, see Fig. 1 (a) and (b)), Scanning and Transmission Electron Microscopy (SEM and TEM) investigations.

## 3. Magnetic properties of ZnCoO films

The magnetic properties of the ALD-grown ZnCoO films were measured in the in-plane configuration (perpendicular to c-hexagonal axis of ZnCoO films) in commercial superconducting quantum interference device (SQUID) magnetometer. Examples of magnetization curves are shown in Fig. 2 (a) and (b) for: (a) sample with the uniform Co distribution and (b) for the one grown at increased temperature with the 2:1 ratio of the ZnO to CoO ALD cycles. The strong non-linear magnetic response at RT (see Fig. 2 (b)) cannot be taken as a signature of the true FM of ZnCoO. We will come back to this issue at the end of the text, here, on the account of the lack of any characteristic features on zero-field cooled and field cooled M(T) which could be clearly attributed to a blocked superparamagnetic behavior (see the bottom-right inset to Fig. 2 (b)), we can rule out Co aggregation into uniformly oriented separated nanocrystals of a narrowly distributed volumes.

We find that RT FR was observed by us only for samples with a sizable nonuniform Co distribution, due to change of the ratio of the ALD cycles, increased growth temperature, or a post-growth annealing (see Fig. 3). In Fig. 3 we collected the data obtained by us at different growth conditions with a growth temperature varied between 160 and 300 $^o$C. Also film thickness varied between 50 and 1400 nm. Thus, it is not surprising that we do not find any correlation between the magnitude of the magnetic moment (correlated to the volume of the ZnCoO layer, say, its 'magnetisation', *M*) and Co concentration (see Fig. 3). The magnetization values are very scattered (see Fig. 3). This statement holds true for a range of samples grown at different ALD sequences, number of cycles (film thickness) or temperatures, however being generally close to those given in paragraph 2.

We then looked for correlations between growth conditions and magnetization values. In Fig. 4 we show relation between the signal strength and sample thickness for ZnCoO films grown with 2:1 ZnO to CoO ALD ratio at 200 $^o$C. This figure indicates that in nonuniform samples RT FR must originate mainly from Co-rich aggregates located either in the near-to-surface layer of the film or in the ZnCoO/Si interface. Most of our samples were grown on silicon, but similar trend, as the one shown in Fig. 4, is found for films grown on other substrates.

We reject here another possible explanation of our data shown in Figs. 2 - 4 that the magnitude of RT FM response relates to a large density of lattice defects (vacancies) at the interface region, as observed recently [7,8]. Firstly, deep defects concentration is very low in our ALD films grown at low temperature. Secondly, even though they are created by a post-growth annealing samples remain paramagnetic if 8 to 1 ratio of the ZnO to CoO ALD cycles was used and samples were grown at temperature below 200 $^{o}$C. Thirdly, for samples with uniform Co (also Mn) distribution paramagnetic response was observed by us independently of a layer thickness. Finally, we never observed FM response in our ALD ZnO samples without Mn or Co doping.

## 4. Origin of RT FM in ZnCoO films grown at low temperature by ALD

Several mechanisms were proposed to account for the RT FR of ZnCoO. Below, we will overview them checking if they can account for the present results.

### 4.1 Defects in ZnO

Several groups claimed that defects such as hydrogen [9] or vacancies [7,8,10] can participate in generation of RT FR in ZnCoO or ZnMnO. We reject the role of Co-H interaction on the ground that we find no influence of H on $M$ until concentration of H reached relatively large values. This effect will be discussed in the forthcoming publication and thus is not discussed here.

Other groups reported that FR was observed in oxygen deficient samples [10,11]. This is why it was claimed that oxygen vacancies are responsible for the FM signal [12]. We doubt, however, that this is the case in our LT ZnCoO films grown by the ALD. Samples grown at LT show only band edge emission and the post-growth annealing is essential to create defects such as vacancies (Fig. 5). Moreover, if generated, we found from depth-profiling cathodoluminescence (CL) investigations (not discussed here) that deep defects are more or less depth uniformly distributed, which cannot account for the observed RT FM properties shown in Fig. 4. Further arguments were given above.

### 4.2 Foreign phases

Formation of different Mn oxides was already observed by us in nonuniform ZnMnO films [5]. Here we report a similar situation for ZnCoO films grown by the ALD. XPS spectra were recorded for all samples studied here, i.e., uniform and nonuniform samples. Here we discuss the results obtained for the sample about 60 nm thick grown at 200 °C with 2:1 ratio of ZnO to CoO ALD cycles. We find a very similar situation in the present study, as the one we observed earlier for ZnMnO films. SEM, Energy Dispersive X-Rays Spectroscopy (EDX), X-Rays Diffraction (XRD) and X-Ray Photoelectron Spectroscopy (XPS, see Fig. 6 (a)) investigations indicate presence of such foreign phases as $Co_2O_3$ and $Co_3O_4$.

We performed depth-profiling XPS investigation to determine depth distribution of these oxides. The depth-profiling was achieved by measuring XPS signals after sequential removing upper layers of the sample by $Ar^+$ sputtering. Details of these investigations will be given elsewhere. Here we only summarise the most important observations. Co oxides were distributed through depth of all samples. No large deviations from the uniformity of their distribution were found. Thus, the Co oxides cannot explain the RT FR in our films. Likely, as in the case of ZnMnO [5], they may contribute to the FM response found at low temperatures, not discussed here. A new XPS signal dominates when data were detected at the interface region. This signal is due to Co metal accumulation occurring at the interface (see Fig. 6 (b)).

**4.3 Co rich regions in ZnCoO**

Their role was proposed recently for ZnCoO films [13]. By Co rich regions we mean here ZnCoO areas with Co concentration above the average one. FM contribution could come here from Co ions at interface of small Co enriched grains. It was claimed that such regions can be of nm sizes and thus difficult to be detected. This is why we looked for micro-visualisation technique which could depict such nonuniformities in Co distribution. For example, we could not find them in TEM investigations performed on our samples.

Recently we proposed a new approach to visualize such TM rich regions in ZnMnO films [14,15]. For this we applied CL method and its depth-profiling option [14,15] to search for TM rich area and their depth distribution. We utilized the fact that Mn ions in ZnO deactivate visible PL. This is also the case for ZnCoO films for which visible emission of ZnO is seen only for films with Co fractions of only a few percent. Thus for the experiment shown in Fig. 7 we selected ZnCoO sample with Co concentration of 3 %. Uniform sample was used (grown with 8:1 ratio of the ALD cycles) and SEM and CL spectra were taken for as-grown sample and after annealing at different temperatures. Only for sample annealed at 800°C CL images start to deviate from the SEM ones, i.e., CL fluctuations are not related to a columnar microstructure of our samples.

Dark regions (see Fig.7) which appeared in the annealed film we relate to regions with increased Co concentration). Depth-profiling indicates that such Co rich regions are in form of columns, rather than small grains. Their distribution is thus depth uniform, which cannot explain the results shown in Fig. 4.

**4.4 Co metal accumulation**

Tendency of Co accumulation in ZnCoO is well documented [16-20]. These accumulations are proved by us to exhibit FM properties. We first searched if such accumulations are present in our films and if so what is their contribution to RT FM.

XPS depth-profiling analysis indicated different chemical compounds and their distribution within the ZnCoO/Si interface region to compare with the surface area. Cobalt oxides were found at the surface region (Fig. 6 (a)). The XPS peaks attributed to Co-metal (Fig. 6 (b)) dominate the spectra collected within the film/substrate interface region and are not detected at the surface (see Fig.6 (a)). As result, one can observe a relative increase of Co concentration within the interface as seen in Fig.6 (b).

We performed RT X-Rays Magnetic Circular Dichroism (XMCD) investigations to verify the postulated here FM origin in our samples. We selected the same sample for which the magnetic data are shown in Fig. 2 and the XPS data are shown in Figs. 6 (a) and (b) (nonuniform 60 nm ZnCoO layer). The thickness variation was obtained upon $Ar^+$ sputtering and thus removing material after the sample was prepared, as reported for the XPS experiments in Section 4.3. A small XMCD signal at the Co $L_{3,2}$ edges under an applied magnetic field of roughly 350 Oe was observed in the interface region. The XMCD response is con consistent with Co in the metallic state. The detail analysis of the XMCD data will be given in the forthcoming publication. This observation decisively points to metallic Co aggregates at the ZnCoO/Si interface as the main source of RT FR in our films. Interestingly, as already stated, the form of *M(T)* precludes simple Co aggregation into magnetic nanocrystals which as an ensemble would exhibit either the superparamagnetic or a blocked-superparamagnetic behavior. The magnetic finding indicates a more complex Co atoms arrangement, perhaps a mesh of magnetically interacting, metallic aggregates.

## 5. Summary

We relate RT FR observed in our ZnO films grown at low temperature by the ALD to the presence of Co metal accumulations at the ZnCoO/Si interface. We reject other possible mechanisms of the FM response in our ZnCoO films, but it should be underlined that they can be important if the origin of low temperature FM is discussed.

**Acknowledgements** The research was partially supported by the EU within the European Regional Development Fund through grant Innovative Economy (POIG.01.01.02-00-008/08) and FunDMS Advanced Grant within the "Ideas" 7th Framework Programme of the EC. The Authors thank prof. Tomasz Dietl for a discussion of the results.

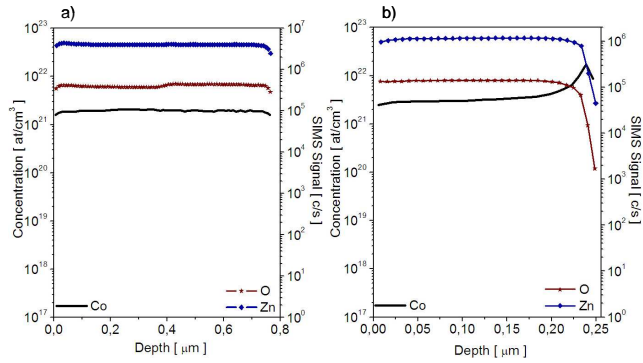

**Figure 1.** (a, b) Depth profile of the ZnCoO film made by SIMS for sample grown at 8 to 1 (a) and 2 to 1 (b) ratios of ZnO to CoO ALD cycles.

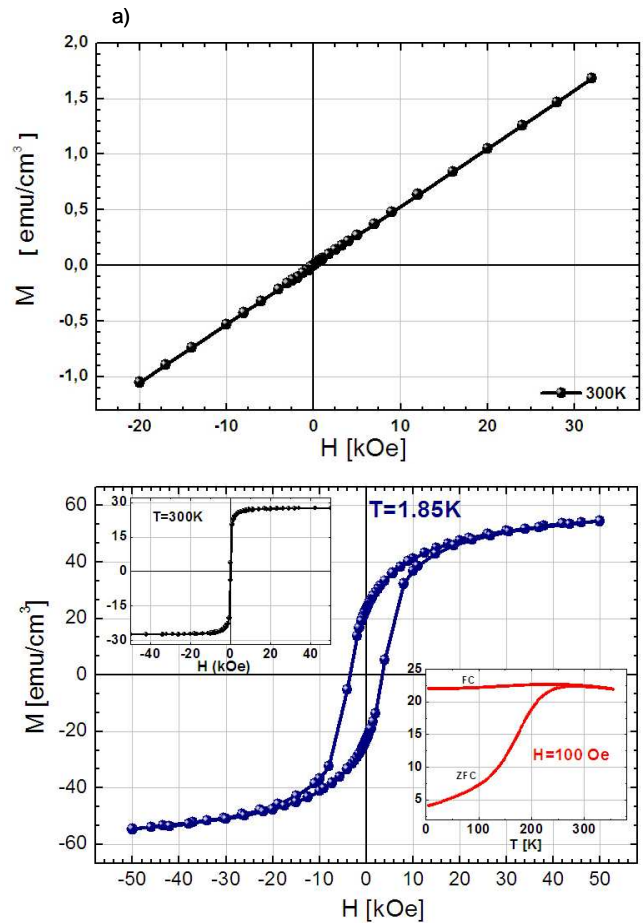

**Figure 2.** Magnetization curves: (a) sample with uniform Co distribution and (b) nonuniform sample grown at increased temperature with the 2:1 ratio of the ZnO to CoO ALD cycles. In the latter case low and RT data are given.

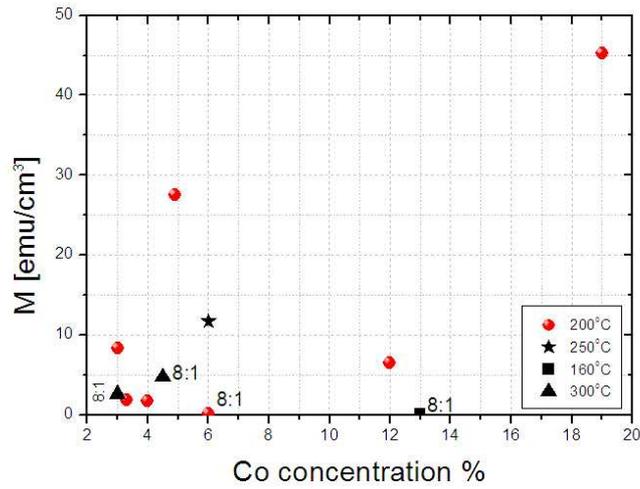

**Figure 3.** Lack of correlation between the magnetization (measured with SQUID) and Co concentration (concentration determined by SIMS and EDX) once data taken for samples grown at different ALD cycles (those with 8:1 ratio are marked) are collected. Growth temperature is indicated.

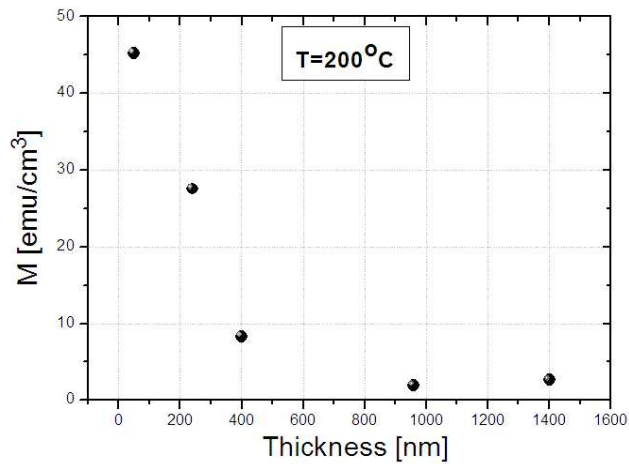

**Figure 4.** Anticorrelation between the magnetization (measured with SQUID) and film thickness. We show results for samples grown with 2:1 ratio of the ZnO to CoO ALD cycles.

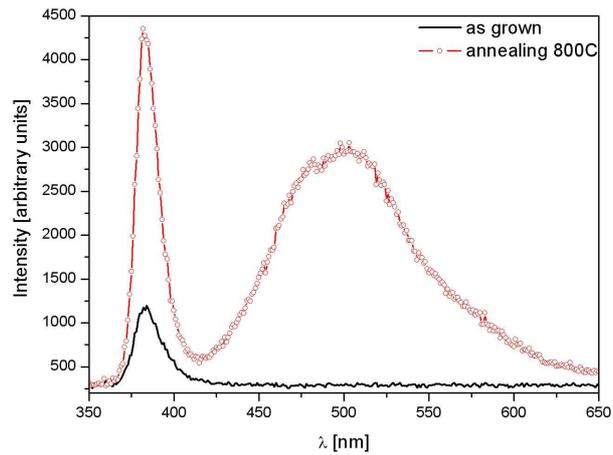

**Figure 5.** Room temperature CL spectra for ZnCoO thin film grown at 160°C. We show the data for as-grown sample and for the same sample annealed at 800 °C.

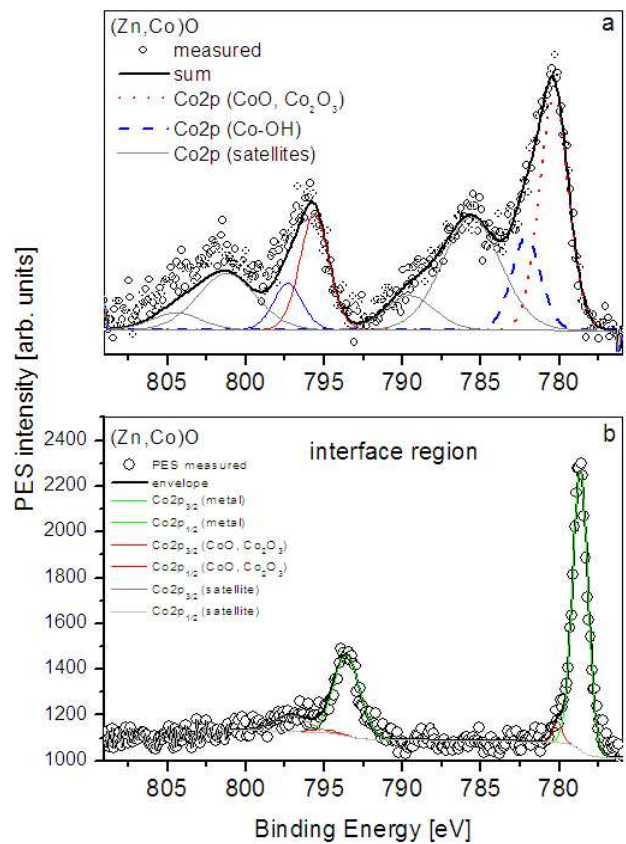

**Figure 6.** The analysis of the Co 2p XPS spectrum taken at the surface region of the ZnCoO/Si sample. Deconvoluted XPS spectra indicate contribution of substitutional Co but also Co oxides and Co-OH. For signal collected from the interface region a new imput from metallic Co is resolved.

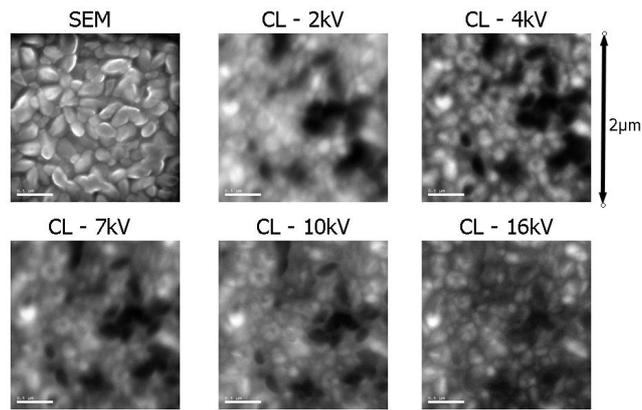

**Figure 7.** CL maps of depth-profiling of 2x2 µm area of ZnCoO film after annealing at 800°C. The data where taken at accelerating voltages from 2 to 16 kV, i.e., for conditions in which signal is collected through all the depth of the sample from its surface (at 2 kV) to the interface region (for V~ 16 kV).